\documentclass[runningheads]{llncs}
\usepackage{graphicx}
\usepackage{multirow}
\usepackage[flushleft]{threeparttable}

\usepackage{xcolor}

\newcommand{\ie}{\textit{i}.\textit{e}.}

\usepackage{amsmath,amsfonts,bm}

\def\eqref#1{equation~\ref{#1}}

\def\1{\bm{1}}

\DeclareMathAlphabet{\mathsfit}{\encodingdefault}{\sfdefault}{m}{sl}
\SetMathAlphabet{\mathsfit}{bold}{\encodingdefault}{\sfdefault}{bx}{n}

\begin{document}
\title{Scribble-based Hierarchical Weakly Supervised Learning for Brain Tumor Segmentation}
\titlerunning{Scribble-based Hierarchical Weakly Supervised Learning}
%

%
\authorrunning{}

\author{
Zhanghexuan Ji\and 
Yan Shen\and 
Chunwei Ma\and  
Mingchen Gao
}
\institute{
Department of Computer Science and Engineering,
University at Buffalo,\\
The State University of New York, Buffalo, USA\\
\email{\{zhanghex,yshen22,chunweim,mgao8\}@buffalo.edu}
}
\maketitle              
\begin{abstract}
The recent state-of-the-art deep learning methods have significantly improved brain tumor segmentation. However, fully supervised training requires a large amount of manually labeled masks, which is highly time-consuming and needs domain expertise. Weakly supervised learning with scribbles provides a good trade-off between model accuracy and the effort of manual labeling. However, for segmenting the hierarchical brain tumor structures, manually labeling scribbles for each substructure could still be demanding. In this paper, we use only two kinds of weak labels, i.e., scribbles on whole tumor and healthy brain tissue, and global labels for the presence of each substructure, to train a deep learning model to segment all the sub-regions.  Specifically, we train two networks in two phases: first, we only use whole tumor scribbles to train a whole tumor (WT) segmentation network, which roughly recovers the WT mask of training data; then we cluster the WT region with the guide of global labels. The rough substructure segmentation from clustering is used as weak labels to train the second network. The dense CRF loss is used to refine the weakly supervised segmentation. We evaluate our approach on the BraTS2017 dataset and achieve competitive WT dice score as well as comparable scores on substructure segmentation compared to an upper bound when trained with fully annotated masks.

\end{abstract}
\section{Introduction}
Image segmentation is a fundamental yet challenging problem in medical imaging domain. Recently, medical image segmentation has been significantly improved by the state-of-the-art fully supervised deep neural networks, e.g., Unet~\cite{cfcn,unet} is widely used in many medical segmentation tasks. However, training such networks requires a large amount of manually annotated masks, which are highly time-consuming, expensive and need domain knowledge.

To reduce the dependence on fully labeled data, semi-supervised and weakly supervised methods have been explored in image segmentation. Semi-supervised learning only needs a small number of fully labeled images and has been studied on tumor segmentation task~\cite{semiae}. A mixed supervision method~\cite{mixsup} is also presented on the same task, where global labels indicating tumor presence are used to instruct the network to extract features from other unlabeled training data. As for weakly supervised learning, scribble-based methods~\cite{recist,scribble,crf} provide a good trade-off between model accuracy and the needs of image annotation. Given the scribbles of each object and background, ScribbleSup~\cite{scribble} applies Graph Cut-based methods to propagate labels to all the image pixels and train a network with these labels. Tang et al.~\cite{crf} directly train network with only the scribble pixels and fine-tune the network with dense CRF loss to refine the object boundaries. In medical imaging, a lesion segmentation network that is trained on RECIST scribbles labeled by doctors with auxiliary bounding boxes as weak labels, shows good segmentation performance on 2D slices~\cite{recist}.

However, current scribble-based methods still require scribbles on every object needs to be segmented, which is not always available especially for medical objects with hierarchical structures. For example, a brain tumor MRI normally has three substructures: edema (ED), enhanced tumor (ET), and non-enhanced tumor (NET/NCR). They form a hierarchical tumor structure: whole tumor (WT) contains all three regions; tumor core (TC) includes ET and NET/NCR; and ET itself~\cite{brats}. In this case, manually scribbling each substructure is still demanding from domain experts, since it requires inspecting different MRI modalities to distinguish the boundaries of substructures~\cite{brats} and adding multiple scribbles carefully in each subregion. By contrast, the scribble markers like RECIST for the whole tumor region and the global labels for the presence of each substructures are relatively easy to acquire in practice. We design the brain tumor substructure segmentation with only the two kinds of aforementioned labels, where the whole tumor can be learned with scribble-based weakly supervised segmentation networks, and global labels are pertinent to guide the clustering of the substructures within the tumor region. Thus, we form a hybrid task with both weakly supervised and unsupervised learning.

Directly extracting the tumor substructures from the whole brain using unsupervised clustering is extremely difficult due to the complicated brain structure and texture. It is highly possible to generate false positives in normal brain regions. In a liver lesion segmentation task, a cascaded model with two networks to hierarchically segment liver and lesion is proposed~\cite{cfcn}.
It would eliminate false positives from areas outside the region of interest. Inspired by that work, we train two networks specifically designed for WT segmentation and substructure segmentation. We first train a network to segment the whole tumor with the given scribble. We recover the whole tumor masks and extract substructures within the tumor region via unsupervised clustering. Finally the second network for substructure segmentation is trained with the clustered labels as weak labels. 

Our contributions can be summarized as: (1) To the best of our knowledge, we are the first to train a model for tumor substructures segmentation with only whole tumor/normal brain scribbles and the global labels. (2) We present a hierarchical weakly supervised learning pipeline for medical structure segmentation which integrates graph-based method, deep learning with CRF loss and unsupervised method. (3) Our results show that our model achieves competitive dice score on WT segmentation and comparable scores on TC and ET segmentation, compared to fully supervised models. 
\section{Methods}
\begin{figure}[t!]
\includegraphics[width=\textwidth,height=0.55\textwidth]{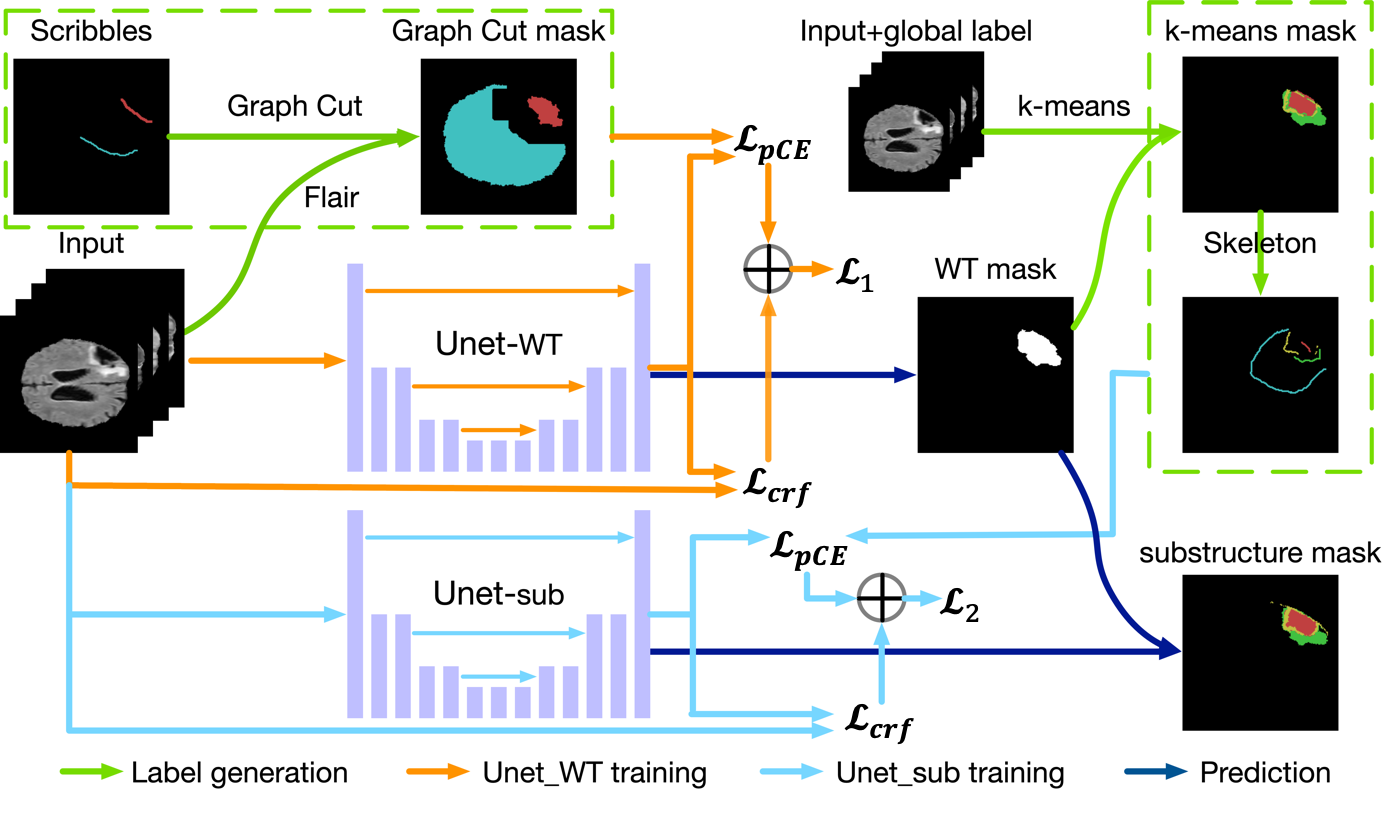}
\caption{Workflow of our proposed method. Label generation forms weak labels for training. Unets are trained on partial cross-entropy loss and finetuned with dense CRF loss.} \label{fig1}
\end{figure}

Our scribble-based hierarchical weakly supervised model for brain tumor segmentation consists of two phases: 1. We first augment the pixel labels from scribbles using Graph Cut and train a Unet called Unet-WT to recover the WT masks of the training scans. 2. We apply k-means clustering with the guidance of global labels within 3D WT region to get the initial substructure segmentation, which is used to train another Unet named Unet-sub to segment TC and ET. Both Unets are first trained on weak labels, and then finetuned with dense CRF loss~\cite{crf}. The pipeline of our approach is visualized in Fig.~\ref{fig1}.

\subsection{Weakly Supervised Whole Tumor Segmentation}
\subsubsection{Initial Tumor Segmentation using Graph Cut.}
We denote the scribble pixels as $S$ and corresponding labels $T_S$. Compared to the number of pixels in MRI scans, the percentage of $S$ is extremely small. Therefore, we apply Graph Cut~\cite{gc} to generate the initial tumor masks and expand the pixels for training. Flair scan is used as the reference image, which contains significant information for whole tumor segmentation~\cite{semiae,hemis}. The 3D Flair MRI is first oversegmented into supervoxels for efficiency. We initialize Graph Cut with WT and normal brain scribbles as seeds, and then finds a min cut via iteratively minimizing the energy function~\cite{gc}. Both foreground and background masks are then eroded to reduce the label ambiguities on the boundaries. The pixels on the boundary are kept unlabeled and will be classified by the following network. We denote the pixels labeled by graph cut as $G$ and labels as $T_G$. Please note that we exclude $S$ from $G$ because they are treated separately in the following weakly supervised learning. 

\subsubsection{Initial Unet-WT Training for Whole Tumor Segmentation.} 
Unet~\cite{unet} is used as our base model due to its efficiency and good performance. We use partial cross-entropy loss for training, which only calculates the loss on the labeled pixels. Scribbles provide ground truth labels, while labels from graph cut may contain error. Based on their credibility, we apply different weights to $S$ and $G$ pixels when calculating the loss. The objective function is defined as: 
\begin{equation}
L_{pCE}=L_S+\lambda L_G=-\frac{1}{|S|}\displaystyle\sum_{i\in S}\displaystyle\sum_{k\in C}t^k_i log(y^k_i) -\lambda \frac{1}{|G|}\displaystyle\sum_{i\in G}\displaystyle\sum_{k\in C}t^k_i log(y^k_i)
\end{equation}
where $i$ for pixel index, $C=\{0,1\}$ is label set, and $k$ for class label in $C$. $y^k_i$ is the probability of pixel $i$ being label $k$ from the network, $t_i$ is the one-hot label vector of pixel $i$ from $T_S$ and $T_G$, and $\lambda$ is the loss balancing weight. 

\subsubsection{Finetuning Unet-WT Using Dense CRF Loss.}
Because of the limited boundary restriction provided in the weak labels, we finetune the network with the dense CRF loss to refine the tumor boundaries~\cite{crf}. The pairwise potential term in CRF energy function can be relaxed to a quadratic form which works for softmax output from the network: $L_{crf}=\sum_{ij}\sum_k \psi(y^k_i,y^k_j)=\sum_{ij}\sum_k y^k_i(1-y^k_j)W_{ij}$, where $i,j$ are pixels, $k$ is label, and $W$ is the affinity matrix between different pixels and is calculated by bilateral Gaussian filter. The finetuning CRF loss is expressed as: 
\begin{equation}
L_{ft}=L_{pCE}+\alpha L_{crf}
\end{equation}
where $\alpha$ is the weight balance parameter. In addition, dense CRF loss, unlike CRF post-processing, refines the boundary prediction during training, which extremely increases the model's inference efficiency. 

\subsection{Weakly Supervised Tumor Substructure Segmentation}
\subsubsection{Initial Substructure Segmentation by K-means Clustering.}
In this step, we need to further segment the WT masks recovered from Unet-WT (denote as $FG_r$) into three substructures: ED, ET, NCR/NET. The challenge is that we have no pixel-level knowledge of any substructures, but only the global label, \ie, which substructures appear in the 3D brain scan. This gives us the motivation to apply unsupervised clustering methods directly on the multimodal 3D MRI scans. Here we simply choose k-means method because of its high efficiency and good performance~\cite{kmeans1}. k is set to be 2 or 3 depending on the global labels.

According to the BRATS benchmark~\cite{brats}, T1ce modality is the main reference for experts to manually label ET and NET/NCR regions, based on the intensity of T1ce and the difference between T1ce and T1. Modality dropping tests~\cite{semiae,hemis} also prove the important role of T1ce for tumor substructure segmentation. Therefore, the feature vector for k-means clustering is constructed as a five-length vector, emphasizing the importance of T1ce and (T1ce-T1). Specifically, $\bm{f}_{i\in FG_r}=[T1_i, 2\cdot T1ce_i, w_{t2}\cdot T2_i, w_{fl}\cdot FL_i, 2\cdot (T1ce-T1)]$, where $i$ denotes each pixel in $FG_r$, and T1, T1ce, T2, FL stand for the normalized intensity of each channel. T1ce and T1ce-T1 are doubled in order to emphasize T1ce for tumor core segmentation. We also set T2 or Flair with smaller weights to prevent them  can dominating $\bm{f}$ and causing bad clustering of tumor core. Therefore, $w_{t2}=min\{1,\frac{s_{t1ce}}{s_{t2}}\}$ and $w_{fl}=min\{1,\frac{s_{t1ce}}{s_{fl}}\}$ are used to reduce the weights of T2 and FL, where $s$ stands for the variance of tumor intensity. 

Since k-means cluster cannot directly give us the labels of each group, we need to classify them by extra appearance and spatial clues~\cite{brats}. The group with highest mean T1ce intensity is labeled as ET, if global label indicates 3 substructures; ED always surrounds the tumor core, so the group with the smallest mean distance to the whole tumor surface is labeled ED, and the remaining group as NET/NCR.

\subsubsection{Training Unet-sub for Tumor Core Segmentation.}
The initial masks of three substructures from k-means clustering are not accurate enough, especially on the boundaries between each class. Image erosion followed by morphological skeletonizing operation are applied to each slice to extract scribbles of each substructure. The scribbles as well as the k-means masks are used as weakly labels to train Unet-sub for the tumor substructures segmentation. The Unet-sub architecture is similar to the Unet-WT architecture, combined loss from scribbles, k-means mask, and dense CRF loss for the refinement. The weight $\lambda$ of k-means loss is set to be smaller than scribbles loss due to its limited accuracy. 

During testing, we merge the ET and NET segmentation output from Unet-sub with the WT mask predicted by Unet-WT. Unet-WT trained on scribbles is able to provide a more accurate WT prediction than Unet-sub. Therefore, the merging process takes WT mask as reference, and translate ET and NET labels from Unet-sub output to the corresponding pixels within the WT mask, and the remaining labels in WT mask are labeled ED. All the pixels excluded from WT mask are labeled as background, even if they might be classified as tumor substructures by Unet-sub. This keeps a high performance of WT segmentation and reduces the false positives from Unet-sub. 

\section{Experiments}
\subsubsection{Dataset Preprocessing.}
We test our method on BRATS2017 dataset~\cite{brats}. It contains 285 brain tumor MRI scans, with four MRI modalities as T1, T1ce, T2, and Flair for each scan. The dataset also provides full masks for brain tumors, with labels for ED, ET, NET/NCR. The segmentation evaluation is based on three tasks: WT, TC and ET segmentation. We randomly separate the dataset into training set with 228 scans and test set with 57 scans. The MRI scans are normalized by subtracting mean and divided by standard deviation, then axial 2D slices are extracted and form 4-channel images as network input. All the slices are center cropped to 192$\times$192 to remove most of the black background. 

To setup our task, we extract foreground and background scribbles from ground truth whole tumor mask. We denote the whole tumor mask as FG, and healthy brain tissues as BG. For FG scribble generation, we select the tumor slices from training set, erode FG with a $3\times3$ kernel and skeletonize the mask with branch cutting so that each mask remains a scribble curve $S$ with 3 pixel width. In order to simulate the randomness of manually scribbles in the real word, we randomly translate each $S$ in both horizontal and vertical orientations by [-20,20] pixels and rotate it by [-30,30] degrees. For BG scribbles, we apply the similar process as above, only this time we first erode BG with a large kernel of $30\times30$. 

\begin{table}[t!]
\caption{Results on BraTS2017 with different settings. Full: training on full (ground truth) masks served as the upper bound of the proposed weakly supervised methods. SC: training on scribbles only. SC+GC: training on both scribbles and Graph Cut masks. CRF: finetuning on dense CRF loss. TrueSC: training on tumor substructure scribbles extracted from the ground truth masks (another upper bound of phase2). KM: training on k-means masks only. KM+SC: training on both k-means masks and scribbles from k-means masks.}\label{tab1}
\centering
\resizebox{\textwidth}{!}{%
\begin{tabular}{l|l|l|l|l|l|l|l}
\hline
\multicolumn{4}{c|}{Phase 1: Unet-WT}                          & \multicolumn{4}{c}{Phase 2: Unet-sub}                                                                 \\ \hline
\multirow{2}{*}{Method} & \multicolumn{3}{c|}{WT}              & \multirow{2}{*}{Method} & \multicolumn{3}{c}{Dice}                  \\ \cline{2-4} \cline{6-8} 
                        & Dice            & Recall & Precision &                         & WT*     & TC              & ET              \\ \hline
Full                    & 0.8987          & 0.9465 & 0.9498       & Full                    & 0.9008 & 0.8063          & 0.8305          \\ \cline{1-4}
SC                      & 0.8487          & 0.9032 & 0.9441       & TrueSC+CRF~\cite{crf}              & 0.8836 & 0.7991          & 0.7920          \\ \cline{5-8} 
SC+CRF                  & 0.8609          & 0.9130 & 0.9461       & KM                      & 0.8516 & 0.6015          & 0.6502          \\
SC+GC                   & 0.8540          & 0.9501 & 0.9030       & KM+SC                    & 0.8643 & 0.6714          & 0.6433          \\
SC+GC+CRF               & \textbf{0.8823} & 0.9443 & 0.9338       & KM+SC+CRF                & 0.8721 & \textbf{0.6988} & \textbf{0.6537} \\ \hline
\end{tabular}%
}
\begin{tablenotes}
  \scriptsize
  \item *WT score in phase2 is the pure WT dice score from Unet-sub. After merging the result from phase1, the final WT dice is the same as phase1.
\end{tablenotes}
\end{table}

\subsubsection{Experimental Implementation Details.}
We select 15224 tumor slices from training set for training. While all the 7729 brain slices from test set are used for testing. The Unet is implemented in Caffe~\cite{caffe}. We use four-channel slices as input to networks. CRF loss is originally implemented for RGB images. Since T1ce has little effect for WT segmentation as well as T2 for TC segmentation~\cite{semiae,hemis}, we use [T1, T2, Flair] channels and [T1, T1ce, Flair] channels to calculate CRF loss when training Unet-WT and Unet-sub respectively. Unet-WT is trained from scratch, and Unet-sub inherits the same parameters as Unet-WT with the last layer changed to four output. Both Unet models are trained for 10k iteration, followed by 20k finetuning on CRF loss, with learning rate linearly reduced from 0.011 to 0, using SGD with momentum 0.9 and batch size of 32. For loss balance weights, $\lambda$ for Unet-WT and Unet-sub training are fixed to $0.8$ and $0.2$ respectively, and $\alpha$ is $10^{-8}$. The proposed method is tested on NVidia 1080ti GPU. We use pixel-level average precision and recall rate, as well as mean Dice coefficient to evaluate the segmentation performance of our method. 

\begin{figure}[t!]
\centering
\includegraphics[width=0.9\textwidth]{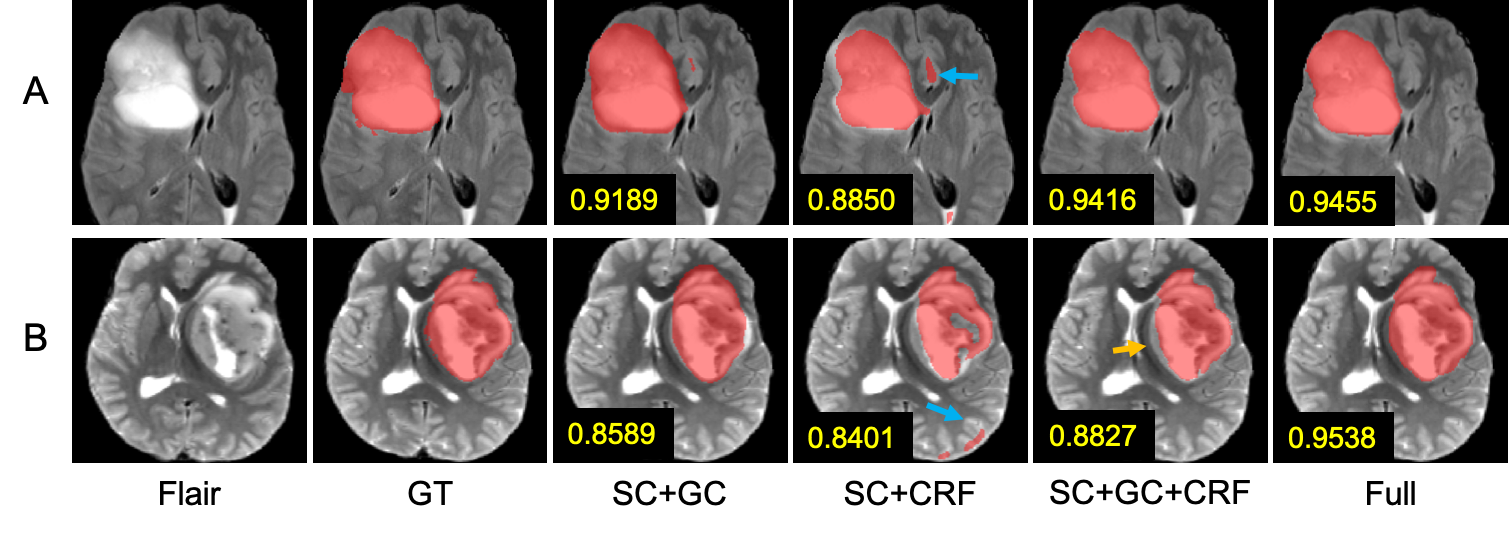}
\caption{Results of two sample slices (A and B) on different settings in phase1. GT: ground truth. Dice scores are reported. Yellow arrow: false negative area. Blue arrow: false positive area.} \label{fig2}
\end{figure}
\begin{figure}[t!]
\centering
\includegraphics[width=\textwidth]{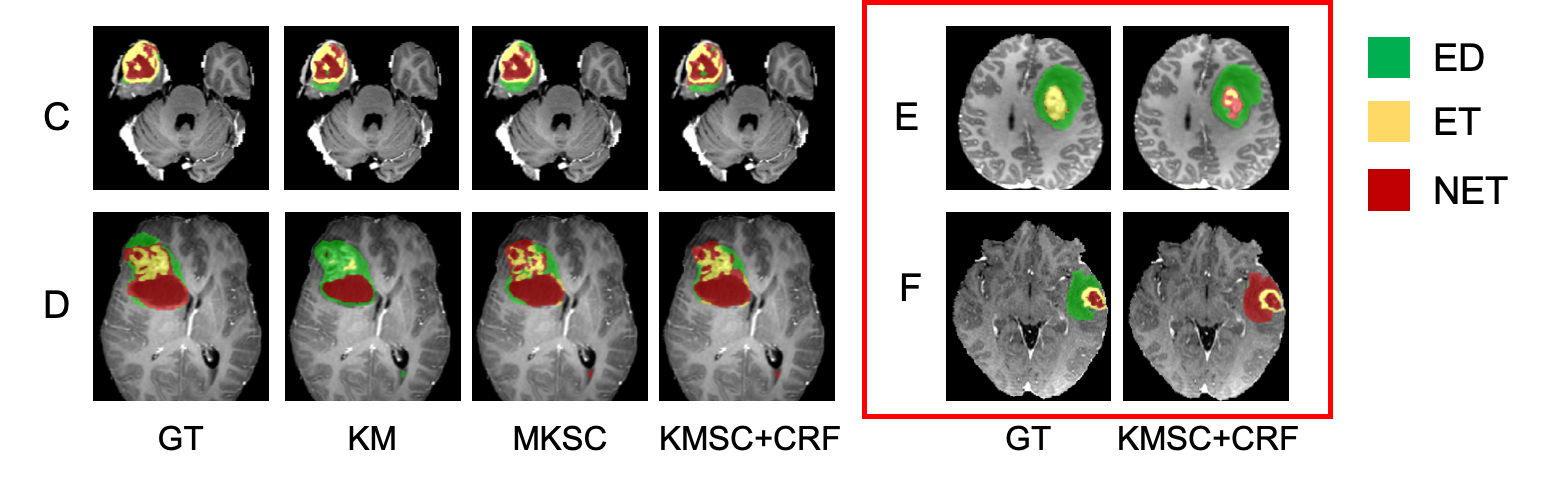}
\caption{Results of another four sample slices (C-F) on different settings in Phase 2. Samples in red box are the mislabeled results.} \label{fig3}
\end{figure}

\subsubsection{Whole Tumor Segmentation Results.} It is shown in Table~\ref{tab1} that our weakly supervised model (SC+GC+CRF) for WT can achieve competitive result that is very close to the fully supervised model with only a 1.64\% margin. Graph Cut is very effective (SC+CRF v.s. SC+GC+CRF) by providing more correct labeled samples to Unet-WT, as shown in Table~\ref{tab1} as well as in Fig.~\ref{fig1}. In addition, finetuning the model with CRF loss efficiently improves the model, which boosts the WT Dice score by $\sim3\%$ compared with the SC+GC setting. As shown in Fig.~\ref{fig2}, CRF loss offers substantial advantages in delineating the tumor boundaries. Fig.~\ref{fig2} shows some negative area in sample B (SC+GC+CRF), that is caused by a strong edge inside the tumor because of the CRF loss.

\subsubsection{Tumor Substructures Segmentation.} 
Our task for Phase 2 is very challenging because there is no pixel-level labels, but only the existence of TC and ET in the training. As shown in Table~\ref{tab1} Phase2, there are still some room for improvement compared with the fully supervised model and the weakly supervised model using scribbles of all the substructures extracted from the GT mask. We still see lots of improvements from our hybrid unsupervised and weakly supervised method compared to the report of other unsupervised methods on brain tumor dataset, which only get Dice score around 60\% for TC and ET~\cite{unsup}. The WT still keeps a high score due to our hierarchical segmentation process. In Fig.~\ref{fig3}, we can see that when assigning high loss weight to the k-means scribbles, the model gives a better performance on TC and ET segmentation in sample C,D, and TC dice score is greatly increased to $\sim70\%$ in Table~\ref{tab1}. The reason is that scribbles contain more correct labels when compared with k-means masks which have unreliable labels on the boundaries. Unsuccessful results are also reported in Fig.~\ref{fig3}, where ED and NET are mislabeled in F, possibly due to their similar feature and appearance in k-means clustering. 

\section{Conclusion}
In this paper we proposed a hierarchically weakly supervised brain tumor segmentation framework with only whole tumor/normal brain scribbles and global labels. Our method achieves competitive WT segmentation performance close to the fully supervised upper bound and comparable TC and ET segmentation. Our methods only require very weak labels, such that has the potential to conveniently make and explore large medical datasets. The unsupervised phase can still be improved by more accurate initializing clustering methods, which is left to our future works. 

\bibliographystyle{splncs04}
\bibliography{paper2319}

\begin{thebibliography}{10}
\providecommand{\url}[1]{\texttt{#1}}
\providecommand{\urlprefix}{URL }
\providecommand{\doi}[1]{https://doi.org/#1}

\bibitem{semiae}
Alex, V., Vaidhya, K., Thirunavukkarasu, S., Kesavadas, C., Krishnamurthi, G.:
  Semisupervised learning using denoising autoencoders for brain lesion
  detection and segmentation. Journal of Medical Imaging  \textbf{4}(4),
  041311 (2017)

\bibitem{recist}
Cai, J., Tang, Y., Lu, L., Harrison, A.P., Yan, K., Xiao, J., Yang, L.,
  Summers, R.M.: Accurate weakly-supervised deep lesion segmentation using
  large-scale clinical annotations: Slice-propagated 3d mask generation from 2d
  recist. In: MICCAI. pp. 396--404. Springer (2018)

\bibitem{cfcn}
Christ, P.F., Elshaer, M.E.A., Ettlinger, F., Tatavarty, S., Bickel, M., Bilic,
  P., Rempfler, M., Armbruster, M., Hofmann, F., D’Anastasi, M., et~al.:
  Automatic liver and lesion segmentation in ct using cascaded fully
  convolutional neural networks and 3d conditional random fields. In: MICCAI.
  pp. 415--423. Springer (2016)

\bibitem{hemis}
Havaei, M., Guizard, N., Chapados, N., Bengio, Y.: Hemis: Hetero-modal image
  segmentation. In: MICCAI. pp. 469--477. Springer (2016)

\bibitem{caffe}
Jia, Y., Shelhamer, E., Donahue, J., Karayev, S., Long, J., Girshick, R.,
  Guadarrama, S., Darrell, T.: Caffe: Convolutional architecture for fast
  feature embedding. In: Proceedings of the 22nd ACM international conference
  on Multimedia. pp. 675--678. ACM (2014)

\bibitem{unsup}
Juan-Albarrac{\'\i}n, J., Fuster-Garcia, E., Manj{\'o}n, J.V., Robles, M.,
  Aparici, F., Mart{\'\i}-Bonmat{\'\i}, L., Garc{\'\i}a-G{\'o}mez, J.M.:
  Automated glioblastoma segmentation based on a multiparametric structured
  unsupervised classification. PLoS One  \textbf{10}(5),  e0125143 (2015)

\bibitem{gc}
Li, Y., Sun, J., Tang, C.K., Shum, H.Y.: Lazy snapping. ACM Transactions on
  Graphics (ToG)  \textbf{23}(3),  303--308 (2004)

\bibitem{scribble}
Lin, D., Dai, J., Jia, J., He, K., Sun, J.: Scribblesup: Scribble-supervised
  convolutional networks for semantic segmentation. In: Proceedings of the IEEE
  CVPR. pp. 3159--3167 (2016)

\bibitem{brats}
Menze, B.H., Jakab, A., Bauer, S., Kalpathy-Cramer, J., Farahani, K., Kirby,
  J., Burren, Y., Porz, N., Slotboom, J., Wiest, R., et~al.: The multimodal
  brain tumor image segmentation benchmark ({BRATS}). IEEE transactions on
  medical imaging  \textbf{34}(10),  1993--2024 (2015)

\bibitem{mixsup}
Mlynarski, P., Delingette, H., Criminisi, A., Ayache, N.: Deep learning with
  mixed supervision for brain tumor segmentation. arXiv preprint
  arXiv:1812.04571  (2018)

\bibitem{unet}
Ronneberger, O., Fischer, P., Brox, T.: U-net: Convolutional networks for
  biomedical image segmentation. In: MICCAI. pp. 234--241. Springer (2015)

\bibitem{crf}
Tang, M., Perazzi, F., Djelouah, A., Ben~Ayed, I., Schroers, C., Boykov, Y.: On
  regularized losses for weakly-supervised cnn segmentation. In: Proceedings of
  ECCV. pp. 507--522 (2018)

\bibitem{kmeans1}
Zhang, L., Gopalakrishnan, V., Lu, L., Summers, R.M., Moss, J., Yao, J.:
  Self-learning to detect and segment cysts in lung ct images without manual
  annotation. In: ISBI 2018. pp. 1100--1103. IEEE (2018)

\end{thebibliography}

\end{document}